\documentclass{PoS}

\title{Searching for intermediate-mass black holes in NGC3310}

\ShortTitle{IMBH candidates in NGC3310}

\author{\speaker{Megan Argo$^{1,2}$}, Joseph Coppola$^1$\thanks{Former MPhys student}\\
        $^{1}$ Jeremiah Horrocks Institute, University of Central Lancashire, Preston PR1 2HE, UK\\
        $^{2}$ Jodrell Bank Centre for Astrophysics, School of Physics and Astronomy, University of Manchester, M13 9PL, UK\\
        E-mail: \email{margo@uclan.ac.uk}}
        
\author{Mar Mezcua$^{3,4}$\\
	$^{3}$ Institute of Space Sciences (ICE, CSIC), Campus UAB, Carrer de Magrans, 08193 Barcelona, Spain\\
	$^{4}$ Institut d'Estudis Espacials de Catalunya (IEEC), Carrer Gran Capit\`{a}, 08034 Barcelona, Spain}
        
\author{Hannah Earshaw$^{5,6}$, Tim Roberts$^5$\\
	$^{5}$ Centre for Extragalactic Astronomy, Department of Physics, Durham University, South Road, Durham, DH1 3LE, UK\\
	$^{6}$ Cahill Center for Astronomy and Astrophysics, California Institute of Technology, Pasadena, CA 91125, USA}

\abstract{Intermediate-mass black holes are theoretically predicted but observationally elusive, and evidence for them is often indirect. The nearby face-on spiral galaxy NGC3310 has hosted many supernovae in recent history, and recent Chandra observations have shown a group of strong off-nuclear X-ray sources that are coincident with radio emission seen in archival VLA and MERLIN observations. Their luminosity, spectrum and off-nuclear location make these sources excellent IMBH candidates. To investigate this possibility, we used combined EVN/e-MERLIN observations at both 1.4 and 5 GHz to look for compact radio emission and evidence of jet activity. We detect a compact radio source within one arcsecond of a Chandra source with an estimated mass ${\rm M}_{\rm BH}\sim3\times10^4 {\rm M}_{\odot}$.}

\FullConference{14th European VLBI Network Symposium \& Users Meeting (EVN 2018)\\
		8-11 October 2018\\
		Granada, Spain}

\begin{document}

\section{Background}
Intermediate-mass black holes (IMBH) are elusive objects, hypothesised to exist in the mass region between stellar-mass black holes (${\rm M_{\rm BH}}<100{\rm M}_{\odot}$) and supermassive black holes (SMBH; ${\rm M_{\rm BH}}>10^{6}{\rm M_{\odot}}$), evidence for which is strong in the centres of galaxies.  IMBHs are potentially the seeds for SMBHs in the early Universe, making them important objects for our understanding of how SMBHs formed and how galaxies have evolved.  Evidence suggests that SMBHs formed soon after the Big Bang (the current record holder is an SMBH of 8$\times$10$^{8}{\rm M_{\odot}}$ recently discovered at a redshift of 7.54, just 690\,Myr after the Big Bang \cite{2018Natur.553..473B}), 
and it is difficult to see how this could happen unless by accretion or the merger/coalescence of IMBHs.  If this is the case, then it is likely that at least some IMBHs did not grow to be SMBHs and have survived to the present epoch.

In the fundamental plane of black hole accretion relating X-ray and radio luminosity and black hole mass (e.g. \cite{2009ApJ...706..404G}) 
there is a distinct gap between the domain of stellar-mass and SMBHs.  While many IMBH candidates exist, observational proof remains frustratingly elusive (see review by 
\cite{2017IJMPD..2630021M}).  Several possible locations for IMBHs have been suggested, and many observational efforts are underway to find them.  One likely location is in the centres of globular clusters where the gravitational potential and crowded environments make coalescence of lower-mass objects to form IMBHs a possibility.  A recent study of the motion of pulsars in 47 Tuc produced evidence for an IMBH at the centre of the cluster \cite{2017Natur.542..203K}, 
although the lack of gas in the cluster centre makes finding a confirmatory electromagnetic signature of accretion an unlikely prospect.  Other likely candidates for IMBHs are off-nuclear Ultra-Luminous X-ray sources (ULXs) with ${\rm L_{X}}>10^{39}$erg/s.  Recent studies 
\cite{2017ARA&A..55..303K} have found that the ULX population is not homogenous; while the majority are likely to be stellar mass black holes or neutron stars accreting in the super-Eddington regime, the possibility remains that the brightest ULX sources with ${\rm L_{X}}\gtrapprox10^{41}$\,erg/s may be active IMBHs (e.g. \cite{2009Natur.460...73F}, \cite{2015MNRAS.448.1893M}). 
 Another promising location for IMBHs is in dwarf galaxies, where many IMBH candidates have been found in the form of low-mass AGN (e.g. \cite{2014AJ....148..136M}, \cite{2016ApJ...817...20M}, \cite{2018MNRAS.478.2576M}, \cite{2018ApJ...863....1C}).  
The detection of gravitational waves is also a potential way of detecting non-electromagentic signatures of IMBH collision and, if they do exist, it is only a matter of time before LIGO detects such an event.

\section{Candidates in NGC3310}
\begin{center}
\begin{figure}[htbp]
   \centering
   \includegraphics[height=7cm]{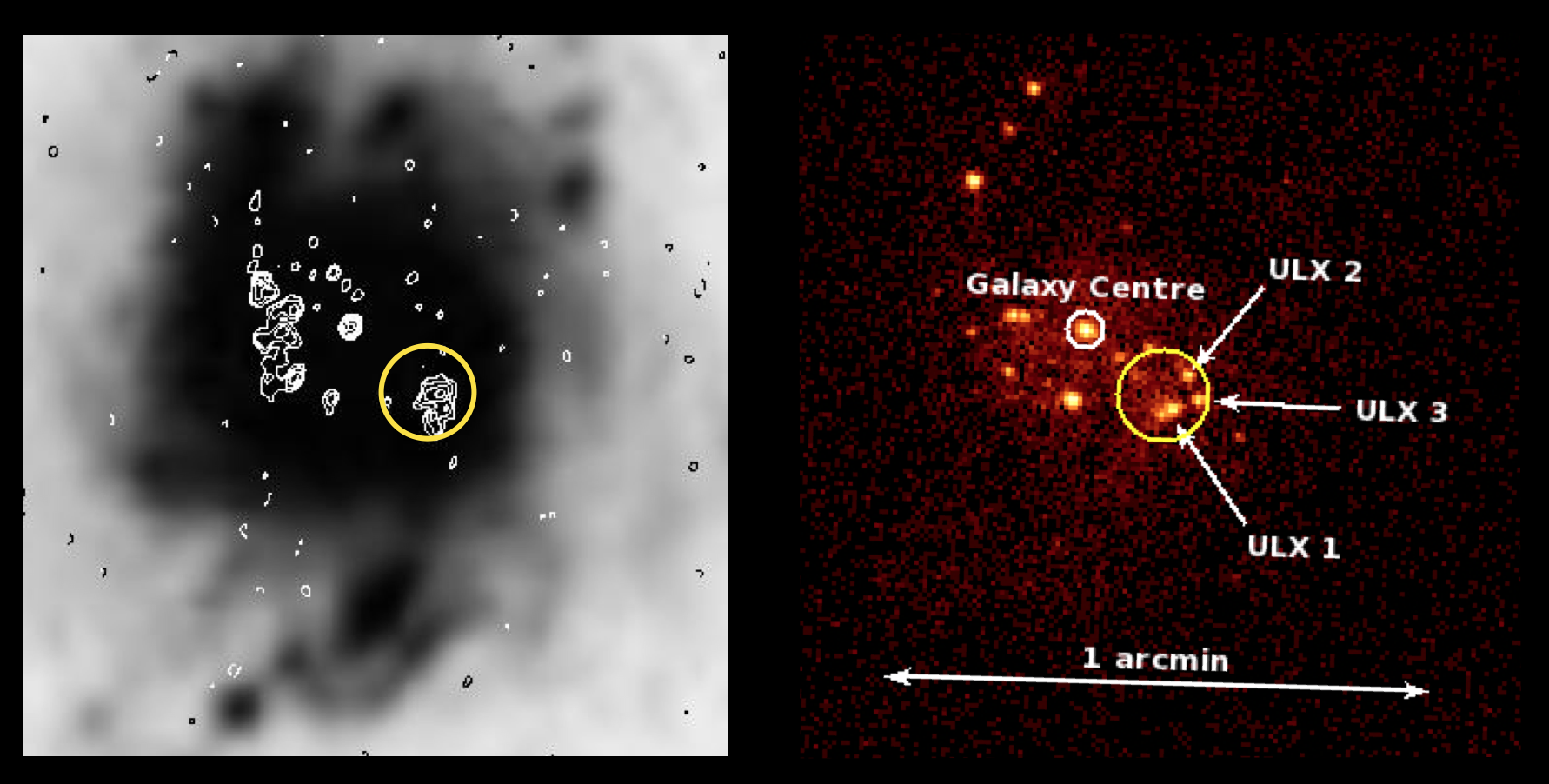}
   \caption{Left: Optical image (greyscale) from the Digitised Sky Survey, with VLA 1.4-GHz radio contours overlaid.  Right: Chandra image of NGC3310 with the positions of the three bright ULX sources in the region of interest.  In both cases, the yellow circle indicates the size of the XMM error circle, centred on the peak of the X-ray emission in the region.}
   \label{fig:comparison}
   
   \vspace{1cm}

   \includegraphics[height=8cm]{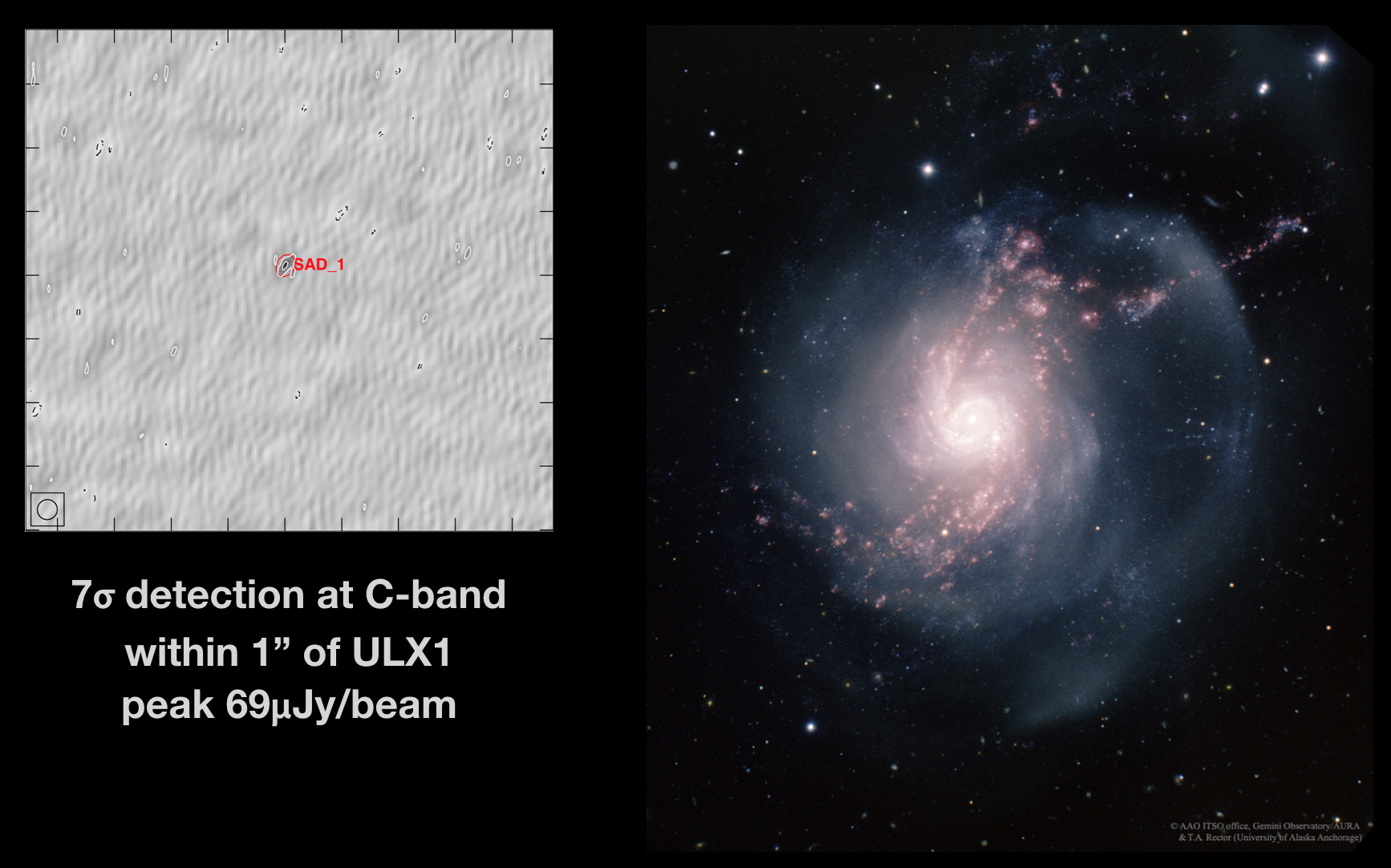}
   \caption{Left: the 7$\sigma$ radio source within one arcsecond of the position of ULX1, as detected at 5GHz with the EVN in 2016.  Right: optical image of NGC3310 showing the disturbed morphology of the outer gas clouds (Gemini).}
   \label{fig:detection}
\end{figure}
\end{center}

There is strong evidence that the face-on spiral galaxy NGC3310 has undergone some kind of close encounter or merger in the past.  It displays significant evidence of past disturbance, a circum-nuclear ring of star formation and evidence of multiple minor merger events (see Figure \ref{fig:detection}).  A number of compact radio sources are known within the central 15 arcsec of NGC 3310 including several historical supernova remnants. 
The galaxy exhibits a high rate of ongoing star formation ($>5M_{\odot} yr^{-1}$). 
Various scenarios to explain NGC3310's morphology have been suggested, including a major merger \cite{2001A&A...376...59K}, 
minor merger \cite{1996ApJ...473L..21S}, 
or multiple minor mergers \cite{2006MNRAS.371.1047W}, 
and it is possible that the cannibalised remains of any interloper may remain in the disk of NGC3310.
Recent simultaneous Chandra/NuSTAR observations have shown that NGC3310 has an overabundance of ULXs thought to be due to the relatively low metallicity of the young stellar population resulting in an excess of luminous X-ray binaries \cite{2015ApJ...806..126L}.


Recent comparison of archival X-ray (XMM/Chandra) and multi-frequency radio (VLA, MERLIN, EVN) data has resulted in the localisation of a radio source with three ULXs in a spiral arm of NGC3310.  Each of these sources has ${\rm L_{X}}>10^{39}$ erg/s, and all are located within a region of radio emission seen at several wavelengths in archival 1.4- and 5-GHz VLA observations (Fig \ref{fig:comparison}).  Their off-nuclear location and coincidence of radio and strong X-ray emission makes these sources excellent candidates for accreting IMBHs.  Despite the large amount of archival data for this galaxy, previous MERLIN observations were not deep enough to detect radio emission from this region (to a 3$\sigma$ noise level of 417 $\mu$Jy/beam), while the VLA data detected radio emission but had insufficient resolution to separate any compact emission from the extended continuum emission in the region.

\section{EVN+e-MERLIN observations}
In order to investigate these candidates further, we obtained EVN+e-MERLIN observations at both L- and C-band in order to look for compact radio counterparts to the X-ray sources, and a simultaneous 35 ks Chandra observation in order to place the source on the fundamental plane of black hole accretion.  Observations were made with the e-MERLIN and EVN arrays simultaneously in October 2016 at L-band (1.6\,GHz) and C-band (5\,GHz). The data consists of two six-hour observations in each band and was correlated at JIVE. The use of both arrays maximises the chance of detecting both compact emission sources as well as extended intermediate-scale jet-like emissions from radio counterparts in the field.  Both observations were split into 8 subbands of 16 MHz each.  NGC3310 had an on-source observation time of ~3.5 hours at both frequencies. DA193 and 3C273 were used as fringe finders while J1044+5322 has used for phase and bandpass calibration. The target was observed in phase-referencing mode with a 2/5-minute calibrator/target cycle. Combining observations from the EVN and e-MERLIN arrays results in a theoretical RMS noise level of ~6$\mu$Jy/beam in each band.

\section{Results}
The EVN C-band data detected a compact 7$\sigma$ source within 1" of the Chandra source ULX1 (Fig \ref{fig:detection}).  Applying the fundamental plane relation of G\"ultekin et al \cite{2009ApJ...706..404G} to the Chandra and EVN detections gives a mass estimate of ${\rm M}_{\rm BH}\sim30,000{\rm M_{\odot}}$. 
We also calculate that log(${\rm R_{X}}) = -4.8$, where ${\rm R_{X} = \nu L_{\nu}(5GHz)/L_{X}(2-10keV)}$ \cite{2003ApJ...583..145T}, 
well within the range expected for IMBHs \cite{2013MNRAS.436.1546M}.  
While the evidence so far suggests an accretion origin for this source, further observational proof is clearly required.  The EVN L-band data unfortunately suffered some telescope failures and the noise level was significantly higher than in the C-band image so at present all we have is an L-band upper limit, limiting our estimates of the spectral index.  We also at present have no variability information for this target.

\section{Further investigations}
ULX1 in NGC3310 is a promising IMBH candidate, but in order to pin down its parameters more reliably we require further observations.  In order to determine the emission mechanism we have requested observations at L- and C-band closely-spaced in time in order to provide spectral index information.  Since IMBHs are compact accreting objects, we also expect them to be variable between flaring and non-flaring states.  The most well-known off-nuclear IMBH candidate to date is HLX-1 in ESO243-49 which shows significant flaring activity on timescales of months in both the radio and X-ray domains (e.g. \cite{2015MNRAS.446.3268C}).  
We have also requested multi-band radio data over more than one epoch in order to investigate the nature of the radio emission from this promising IMBH candidate in NGC3310.  In order to place the object on the fundamental plane of black hole accretion, we are also seeking to obtain a simultaneous X-ray flux.



\bibliographystyle{unsrt}
\bibliography{bibliography}{}


\end{document}